\documentclass[10pt,twocolumn]{article}          
\usepackage{latex8}                              
\usepackage{times}                               
\usepackage{graphicx}

\begin{document}

\title{Parallel netCDF: A Scientific High-Performance I/O Interface}


\author{
  Jianwei Li~~~~~~Wei-keng Liao~~~~~~Alok Choudhary \\
  {\em ECE Department, Northwestern University} \\
  \{jianwei, wkliao, choudhar\}@ece.northwestern.edu \\
  \\
  Robert Ross~~~~~~Rajeev Thakur~~~~~~William Gropp \\
  {\em MCS Division, Argonne National Laboratory} \\
  \{rross, thakur, gropp\}@mcs.anl.gov
}

\maketitle

\begin{abstract}

Dataset storage, exchange, and access play a critical role in scientific applications. For such
purposes netCDF serves as a portable and efficient file format and programming interface, which is popular in 
numerous scientific application domains. However, the original interface does not provide a
efficient mechanism for parallel data storage and access.

In this work, we present a new parallel interface for writing and reading netCDF datasets. This
interface is derived with minimum changes from the serial netCDF interface but defines semantics for 
parallel access and is tailored for high performance. The underlying parallel I/O is achieved through MPI-IO,
allowing for dramatic performance gains through the use of collective I/O optimizations. 
We compare the implementation strategies with HDF5 and analyze both.
Our tests indicate programming convenience and
significant I/O performance improvement with this parallel netCDF interface.

\end{abstract}

\section{Introduction}

Scientists have recognized the importance of portable and
efficient mechanisms for storing large datasets created and used
by their applications. The Network Common Data Form (netCDF)
\cite{ReDa90, RDED97} is one such mechanism used by a number of
applications.

NetCDF intends to provide a common data access method for atmospheric science applications
to deal with a variety of data types that encompass single-point observations, time series,
regularly spaced grids, and satellite or radar images \cite{RDED97}. Today several organizations
have adopted netCDF as a data access standard \cite{WINU}.

The netCDF design consists of both a portable file format and an
easy-to-use application programming interface (API) for
storing and retrieving netCDF files across multiple platforms.
More and more scientific applications choose netCDF as their output file format.
While these applications become computational and data intensive, they
tend to be parallelized on high-performance computers.
Hence, it is highly desirable to have an efficient parallel programming interface to the netCDF files.
Unfortunately, the original design of netCDF interface is proving
inadequate for parallel applications because of its lack of
a parallel access mechanism. In particular, there is no
support for concurrently writing to a netCDF file. 
Hence, parallel applications operating on netCDF files must
serialize access. Traditionally, parallel applications write to netCDF files 
through one of the allocated process which easily becomes a performance bottleneck.
The serial I/O access is both slow and cumbersome
to the application programmer.

To facilitate parallel I/O operations, we have defined a parallel API for concurrently accessing netCDF
files. With minimum changes to the names and argument lists, this interface maintains the look
and feel of the serial netCDF interface while the implementation underneath incorporates well-known
parallel I/O techniques such as collective I/O to allow high-performance data access. 
We implement this work on top of MPI-IO, which is
specified by MPI-2 standard \cite{GrLT99, Mess97, GLDS96} and is freely available on most platforms.
Since MPI has become de facto parallel mechanism for communication and I/O on most parallel environments, 
this approach is portable across different platforms.

Hierarchical Data Format version 5 (HDF5) \cite{HDF5} also provides a portable file format and 
programming interfaces for storing multidimensional arrays together with ancillary data in a single file.
It supports parallel I/O and its implementation is also built on top of MPI-IO. 
However, the HDF5 API is too flexible and cumbrous to become an easy-to-use standard
since it adds more programming features and completely re-designs the API from its previous version. 
Our parallel netCDF interface, on the other hand, is more concise, closer to the original API,
and goes much closer to MPI-IO interface, which introduces less overhead while
providing more optimization opportunities for performance enhancement. The goal of this work is to make the
parallel netCDF interface a data access standard for parallel scientific applications.

We run a couple of benchmarks using parallel netCDF and parallel HDF5, exploring both artificially
made access patterns from our own benchmark and the ones from a real application called FLASH \cite{FORT00}. 
In our experiments, parallel netCDF brings significant I/O improvement and shows better performance 
than parallel HDF5 in the FLASH I/O benchmark \cite{FLASHIO}. 

The rest of this paper is organized as follows. Section \ref{sec:related} reviews some related work.
Section \ref{sec:background} presents the design background of netCDF and points out its potential usage 
in parallel scientific applications. The design and implementation of our parallel netCDF 
is described in Section \ref{sec:design}.
Experimental performance results are given in Section \ref{sec:results}. 
Section \ref{sec:conclusion} concludes the paper.

\section{Related Work}
\label{sec:related}

Considerable research has been done on data access for scientific applications. The work has focused 
on data I/O performance and data management convenience. Two projects, MPI-IO and HDF,
are most closely related to our research.

MPI-IO is a parallel I/O interface specified in the MPI-2 standard. It is implemented and used on a
wide range of platforms. The most popular implementation, ROMIO 
\cite{TRLG02} is implemented portably on top of an
abstract I/O device layer \cite{ThGL96,ThGL99b} that enables portability to new underlying
I/O systems. One of the most important features in ROMIO is collective I/O operations, which
adopt a two-phase I/O strategy \cite{RoBC93,TBCP94,ThCh96,ThGL99} and improve the parallel I/O
performance by significantly reducing the number of I/O requests that would otherwise result in
many small, noncontiguous I/O requests. However, MPI-IO reads and writes 
data in a raw format without providing any functionality to effectively manage the associated metadata. 
Nor does it guarantee data portability, thereby making it inconvenient for scientists 
to organize, transfer, and share their application data.

HDF is a file format and software, developed at NCSA, for storing, retrieving, analyzing, visualizing, and converting scientific
data. The most popular versions of HDF are HDF4 \cite{HDF4} and HDF5 \cite{HDF5}. 
The design goal of HDF4 is mainly to deal with sequential data access and its APIs are consistent with its 
earlier versions. On the other hand, HDF5 is a major revision in which its APIs are completely re-designed.
Both versions store multidimensional arrays together with ancillary data 
in portable, self-describing file formats.
The support for parallel data access in HDF5 is built on top of MPI-IO, which ensures its portability
since MPI-IO has become a de facto standard for parallel I/O.
However, the fact that HDF5 file format is not compatible with HDF4 can be inconvenient for existing HDF4
programmers to migrate their applications to HDF5. 
Furthermore, HDF5 adds several new features, such as hierarchical file structure, to describe more metadata,
but it also increases the difficulties for the implementation of parallel data access underneath.
And the overhead involved may make HDF5 perform much worse than its underlying MPI-IO.
By using a number of scientific applications, this problem is addressed in \cite{LLCT02, RNCZ01}.

\section{NetCDF Background}
\label{sec:background}

NetCDF is an abstraction that supports a view of data as a collection of self-describing, portable,
array-oriented objects that can be accessed through a simple interface. It defines a file
format as well as a set of programming interfaces for storing and retrieving
data in the form of arrays in netCDF files. We first describe the netCDF file format and its
serial API and then consider various approaches to access netCDF files in
parallel computing environments.

\subsection{File Format}

NetCDF stores data in an array-oriented dataset, which contains dimensions, variables, and
attributes. Physically, the dataset file is divided into two parts: file header and array data. The
header contains all information (or metadata) about dimensions, attributes, and variables except
for the variable data itself, while the data part contains arrays of variable values (or raw data).

The netCDF file header first defines a number of dimensions, each with a name and a length. These
dimensions are used to define the shapes of variables in the dataset. One dimension can be
unlimited and is used as the most significant dimension (record dimension) for growing-size
variables.

Following the dimensions, a list of named attributes are used to describe the properties of the
dataset (e.g., data range, purpose, associated applications ). These are called global
attributes and are separate from attributes associated with individual variables.

The basic units of named data in a netCDF dataset are variables, which are multidimensional
arrays. The header part describes each variable by its name, shape, named attributes, data type,
array size, and data offset, while the data part stores the array values for one variable after
another, in their defined order.

\begin{figure}
\begin{center}
\includegraphics[width=0.40\textwidth]{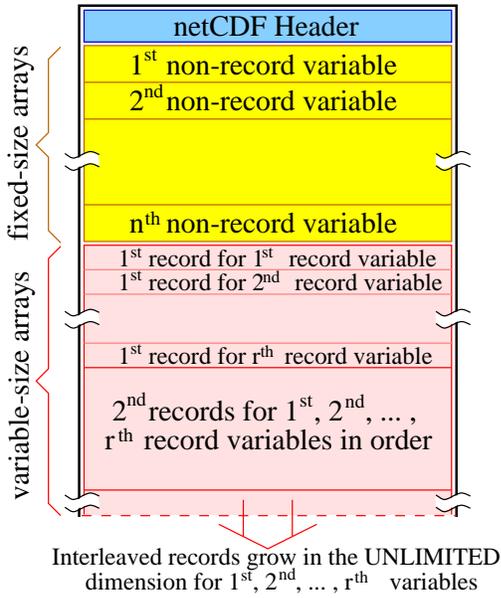}
\end{center}
\vskip -0.1in \caption{NetCDF file structure: there is a file header containing
metadata of the stored arrays, then the fixed-size arrays are laid out in the 
following contiguous file space in a linear order, with variable-size arrays 
appending at the end of the file in an interleaved pattern.
} 
\label{figure:fileformat}
\end{figure}

To support variable-size arrays (e.g., data growing with
time stamps), netCDF introduces record variables and uses a special
technique to store such data. All record variables share
the same unlimited dimension as their most significant dimension
and are expected to grow together along that dimension. The rest,
less significant dimensions all together define the shape for one
record of the variable. For fixed-size array, each array is stored in
a contiguous file space starting from a given offset. For variable-size
arrays, netCDF first defines a {\it record} of an array as a subarray 
comprising all fixed dimensions and the records of all such arrays are 
stored interleaved in the arrays' defined order.
Figure~\ref{figure:fileformat} illustrates the storage layouts for fixed-size
and variable-size arrays in a netCDF file.

In order to achieve network transparency (machine-independence),
both the header and data parts of the file are represented in an
well-defined format similar to XDR (eXternal Data
Representation) but extended to support efficient storage of
arrays of non-byte data.

\subsection{Serial NetCDF API}

The original netCDF API was designed for serial codes to perform
netCDF operations through a single process. 
In the serial netCDF library, a typical sequence of operations to write a new netCDF dataset is to
create the dataset; define the dimensions, variables and attributes; write variable data; and close
the dataset. Reading an existing netCDF dataset involves first opening the dataset; inquiring about
dimensions, variables, and attributes; reading variable data; and closing the dataset.

These netCDF operations can be divided into the following five categories. Refer to
\cite{RDED97} for details of each function in the netCDF library.

\newcounter{Lcount}

\begin{list}{(\arabic{Lcount})}{\usecounter{Lcount}\setlength{\rightmargin}{\leftmargin}}
\item \textbf{Dataset Functions}: create/open/close a dataset, set the dataset to define/data mode, and synchronize dataset
changes to disk
\item \textbf{Define Mode Functions}: define dataset dimensions and variables
\item \textbf{Attribute Functions}: manage adding, changing, and reading attributes of datasets
\item \textbf{Inquiry Functions}: return dataset metadata: dim(id, name, len), var(name, ndims, shape, id)
\item \textbf{Data Access Functions}: provide the ability to read/write variable data in one of the five access methods: single value, whole array,
subarray, subsampled array (strided subarray) and mapped strided subarray
\end{list}

The I/O implementation of the serial netCDF API is built on the native I/O system calls and has its
own buffering mechanism in user space. Its design and optimization techniques are suitable for
serial access but are not efficient or even not possible for parallel access, nor do they allow
further performance gains provided by modern parallel I/O techniques.

\subsection{Using NetCDF in Parallel Environments}

\begin{figure*}
\begin{center}
\includegraphics[width=0.98\textwidth]{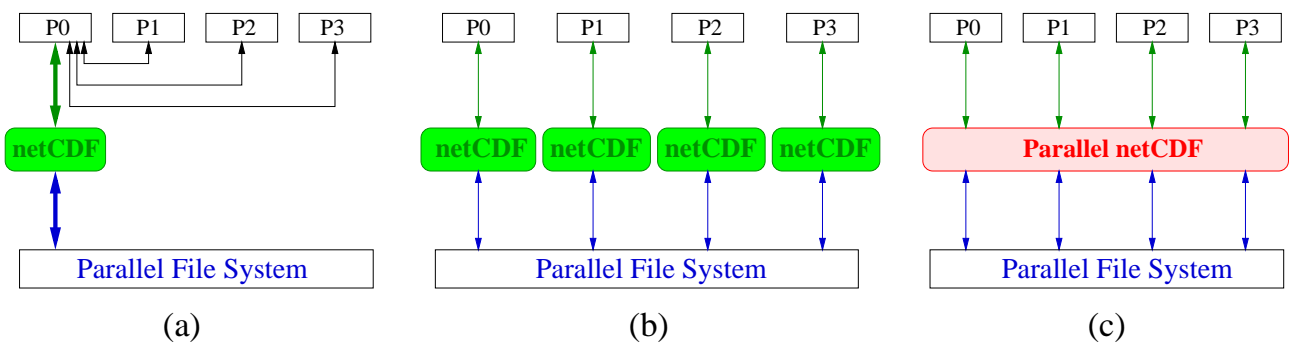}
\end{center}
\vskip -0.1in \caption{Using netCDF in parallel programs:
(a) use serial netCDF API to access single files through a single process;
(b) use serial netCDF API to access multiple files concurrently and independently;
(c) use new parallel netCDF API to access single files cooperatively or collectively. 
}
\label{figure:parallelization}
\end{figure*}

Today most scientific applications are programmed to run in parallel environments
due to the increasing requirements on data amount and computational resources. It is highly desirable
to develop a set of parallel APIs for accessing netCDF files that employs appropriate parallel I/O 
techniques. In the meantime, programming convenience is also important, since scientific users may 
desire to spend minimum effort on dealing with I/O operations.
Before presenting our design on parallel netCDF, we would like to discuss current approaches for 
using netCDF in parallel programs in a message-passing environment.

The first and most straightforward approach is described in the
scenario of Figure~\ref{figure:parallelization}(a) in which one process is in charge of
collecting/distributing data and performing I/O to a single netCDF file using the serial netCDF API. 
The I/O requests from other processes are carried out by shipping all the data through this single process. 
The drawback of this approach is that collecting all I/O data on a single process 
can easily cause an I/O performance bottleneck and may overwhelm its memory capacity.

To avoid unnecessary data shipping, an alternative approach is
to have all processes perform their I/O independently using the serial
netCDF API, as shown in Figure~\ref{figure:parallelization}(b). In this case,
all netCDF operations can proceed concurrently, but over multiple files, one for each process. 
However, using multiple files to store a single netCDF dataset results the complexity and 
difficulty of data management. This approach also destructs the purpose of netCDF design on
easy data integration and management.

A third approach introduces a new set of APIs with parallel access semantics and optimized parallel I/O
implementation such that all processes perform I/O operations cooperatively or collectively through the
parallel netCDF library to access a single netCDF file. This approach, 
as shown in Figure~\ref{figure:parallelization}(c), both frees the users from
dealing with details of parallel I/O and provides more opportunities for 
employing various parallel I/O optimizations in order to obtain higher performance. We discuss the details
of this parallel netCDF design and implementation in the next section.

\section{Parallel NetCDF}
\label{sec:design}

To facilitate convenient and high-performance parallel access to netCDF files, we define a
new parallel interface and provide a prototype implementation.
Since a large number of existing users are running their applications over netCDF, our
parallel netCDF design retains the original netCDF file format (version 3) and introduces minimum
changes from the original interface.  We distinguish the parallel API
from the original serial API by prefixing the C function calls with ``ncmpi\_'' and the
Fortran function calls with ``nfmpi\_''.

\subsection{Interface Design}

\begin{figure}
\begin{center}
\includegraphics[width=0.45\textwidth]{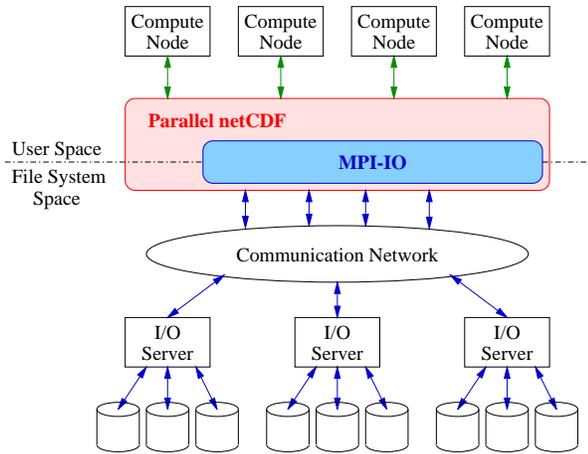}
\end{center}
\vskip -0.1in \caption{Design of parallel netCDF on a parallel I/O architecture.
Parallel netCDF runs as a library between user space and file system space.
It processes parallel netCDF requests from user compute nodes and, after optimization,
passes the parallel I/O requests down to MPI-IO library, and then the I/O servers receive the 
MPI-IO requests and perform I/O over the end storage on behalf of the user. 
}
\label{figure:hierarchy}
\end{figure}

Our parallel netCDF API is built on top of MPI-IO.  
The parallel netCDF built on MPI-IO can benefit from several well-known optimizations
already used in existing MPI-IO implementations,
such as data sieving and two-phase I/O strategies \cite{RoBC93, TBCP94, ThCh96,ThGL99} in ROMIO.
Figure~\ref{figure:hierarchy} describes the overall architecture for our design.

In parallel netCDF, a file is opened, operated, and closed by
the participating processes in a communication group. In order for these
processes to operate on the same file space, especially the structural
information contained in the file header, a number of changes have been
made to the original serial netCDF API.

For the function calls that create/open a netCDF file, an MPI communicator is 
added in the argument list to define the participating I/O processes within the file's open and close scope.
MPI\_Info object is also added to pass user access hints to the MPI-IO for further optimizations. 
By describing the collection of
processes with a communicator, we provide the underlying implementation with information that
can be used to ensure file consistency. The MPI\_Info hint provides users the ability to
deliver the high level access information to netCDF and MPI-IO libraries, such as file access patterns 
and file system specifics to direct optimization.

We keep the same syntax and semantics for the parallel netCDF 
define mode functions, attribute functions, and inquiry
functions as the original ones.
These functions are also made collective to guarantee consistency of dataset
structure among the participating processes in the same MPI communication group. 
For instance, the define mode functions is required to be called by
the processes with the same values.

The major effort of this work is the parallelization of the data access functions.
We provide two sets of data access APIs: a {\it high-level API} that mimics the serial netCDF
data access functions and serves an easy path for original netCDF users to migrate to the parallel
interface, and a {\it flexible API} that provides a more MPI-like style of access. Specifically, the
flexible API uses more MPI functionality in order to provide better handling of internal data representations
and to more fully expose the capabilities of MPI-IO to the application programmer.
The major difference between the two is the use of MPI derived data types.
We believe using MPI derived datatypes can better illustrate the access patterns than the subarray
mapping methods used in original API.

\begin{figure}
\begin{center}
\includegraphics[width=0.45\textwidth]{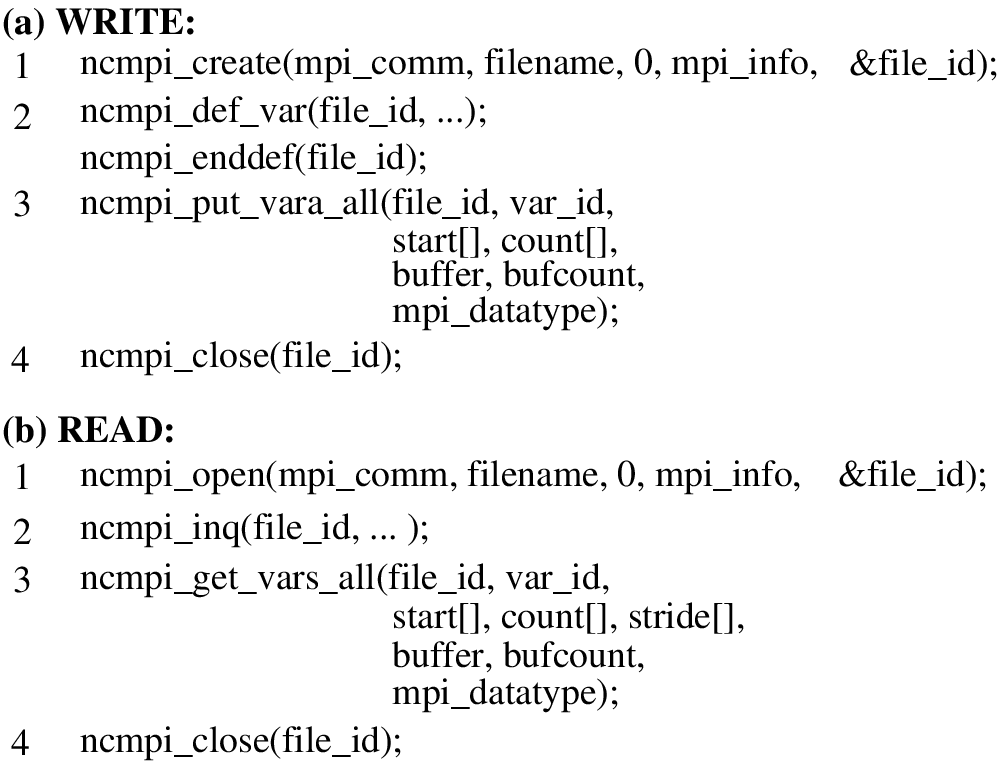}
\end{center}
\vskip -0.1in \caption{Example of using parallel netCDF. Typically there are 4 main steps: 
1. collectively create/open the dataset; 
2. collectively define the dataset by adding dimensions, variables and attributes in WRITE, or
inquiry about the dataset to get metadata associated with the dataset in READ;
3. access the data arrays (collective or non-collective);
4. collectively close the dataset.
}
\label{figure:code}
\end{figure}

The most important change from the original netCDF interface with respect to data access functions is
the split of data mode into two distinct modes: collective and non-collective data modes
in which collective function names end with ``\_all''. Similar to MPI-IO, the collective functions are 
synchronous across the processes in the communicator associated to the opened netCDF file, 
while the non-collective functions are not. Using collective operations 
can provide the underlying parallel netCDF implementation an
opportunity to further optimize access to the netCDF file. These optimizations are performed
without further intervention by the application programmer and have been proven to provide dramatic  
performance improvement in multidimensional dataset access \cite{ThGL99}. Figure~\ref{figure:code} gives an
example code of using our parallel netCDF API to write and read a dataset using collective I/O.

\subsection{Parallel Implementation}

The parallel API implementation is discussed in two parts: header I/O and parallel data I/O.
We first describe out implementation strategies for 
dataset functions, define mode functions, attribute functions, and inquiry functions that access the
netCDF file header.

\subsubsection{Access to File Header}

Internally, the header is read/written only by a single process, although a copy is cached in local
memory on each process. The define mode functions, attribute functions, and inquiry functions all
work on the local copy of the file header. Since they are all in-memory operations not involved
in any file I/O, they bear few changes from the serial netCDF API. They are made collective, but this feature 
does not necessarily imply inter-process synchronization. In some cases, however, when the
header definition is changed synchronization is needed to verify that
the values passed in by all processes match. In all possible cases we
allow inter-process communications.

The dataset functions, unlike the other functions cited, need complete reimplementation because they are in charge of
collectively opening/creating datasets, performing header I/O and file synchronization for all
processes, and managing inter-process communication. We build these functions over MPI-IO so that
they have better portability and provide more optimization opportunities. The basic idea is to let
the ROOT process fetch the file header, broadcast it to all processes when opening a file, and 
write the file header at the end of definition if any modification occurs in
the header part. Since all define mode and attribute functions are collective and require all processes
in the communicator to provide the same arguments when adding/removing/changing definitions, the
local copies of the file header shall be the same across all processes once the file is
collectively opened and until it is closed.

\subsubsection{Parallel I/O for Array Data}

Since the majority of time spent accessing a netCDF file is in data access,
the data I/O must be efficient. By implementing the data access functions above
MPI-IO, we enable a number of advantages and optimizations.

For each of the five data access methods in the flexible data access functions, we represent the
data access pattern as an MPI file view (a set of data visible and accessible from an open file
\cite{Mess97}), which is constructed from the variable metadata (shape, size, offset, etc.) in the
netCDF file header and start[], count[], stride[], imap[], mpi\_datatype arguments provided by
users. For parallel access, particularly for collective access, each process has a different file
view; and all processes in combination can make a single MPI-IO request to transfer large
contiguous data as a whole, thereby preserving useful semantic information that would otherwise be
lost if the transfer were expressed as per process noncontiguous requests.

The high-level data access functions are implemented in terms of the flexible data access
functions, so that existing users migrating from serial netCDF can also benefit from the MPI-IO
optimizations. However, the flexible data access functions are closer to MPI-IO and hence incur less
overhead. They accept a user-specified MPI derived datatype and pass it directly to MPI-IO for
optimal handling of in-memory data access patterns.

In some cases (for instance, in record variable access) the data is stored interleaved by record and the
contiguity information is lost, so the existing MPI-IO collective I/O optimization may not help. In
that case, we need more optimization information from users, such as the number, order, and record
indices of the record variables they will access consecutively. With such information we can collect multiple
I/O requests over a number of record variables and optimize the file I/O over a large pool of data
transfers, thereby producing more contiguous and larger transfers. This kind of information is
passed in as an MPI\_Info hint when a user opens or creates a netCDF dataset. We implement our
user hints in parallel netCDF for all such specific optimization points, while a number of standard
hints are passed down for MPI-IO to take control of optimal parallel I/O behaviors. Thus
experienced users have the opportunity to tune their applications for further performance
gains.

\subsection{Advantages and Disadvantages}

There are a number of advantages within the design and implementation of our parallel netCDF, as compared
to other related work, like HDF5.

First of all, the parallel netCDF design and implementation is optimized for the netCDF file format so that
the data I/O performance is as good as the MPI-IO.
The NetCDF file chooses linear data layout, in which the data arrays are either stored in contiguous space and
in a predefined order or interleaved in a regular pattern. This regular and highly predictable data layout
enables the parallel netCDF data I/O implementation to simply pass the data buffer, metadata (fileview,
mpi\_datatype, etc.), and other optimization information to MPI-IO, and all parallel I/O
operations are carried out in the same manner as when MPI-IO alone is used.  Thus, there is very little
overhead, and the parallel netCDF performance should be nearly the same as MPI-IO if only raw data
I/O performance is compared. On the other hand, parallel HDF5 uses tree-like file structure that are
similar to the UNIX file system and the data is dispersedly laid out using super block, header blocks, 
data blocks, extended header blocks and extended data blocks. This irregular layout pattern may make it difficult
to pass user access pattern directly to MPI-IO especially for the case of variable-size arrays. 
Instead, parallel HDF5 uses dataspace and hyperslabs to define the data organization, 
map and transfer data between memory space and the file space and 
does buffer packing/unpacking in a recursive way, 
while these can otherwise be directly handled by MPI-IO in a more efficient and optimized way.

Secondly, the parallel netCDF implementation manages to keep the overhead involved in header I/O as low as 
possible. In the netCDF file, there is only one header which contains all necessary information for direct
access of each data array and each array is associated with a predefined, numerical ID that can be efficiently 
inquired when it is needed to access the array. So, by maintaining a local copy of the header on each process, 
our implementation saves a lot of inter-process synchronization as well as avoids repeated access of the 
file header each time the header information is needed to access a single array. All header information can
be accessed directly in local memory and inter-process synchronization is needed only during the definition
of the dataset. And once the definition of the dataset is created, each array can be identified by its
permanent ID and accessed at any time by any process, without any collective open/close operation.
However, in HDF5, the header metadata is dispersed in separate header blocks for each object
and, in order to operate on an object, it has to iterate through the entire namespace to get the header information
of that object and then open, access and close it. This kind of access method may be inefficient for parallel
access, since the parallel HDF5 designs the open/close of each object as collective operations, which force
all participating processes to communicate when accessing one single object, not to mention the cost of 
file access to locate and fetch the header information of that object.

Lastly, the programming interface of the parallel netCDF is concise and designed for easy usage, 
and the file format is fully compatible with serial netCDF. 
Porting existing serial netCDF application to parallel netCDF should be straightforward 
because the parallel API contains nearly all functions of the serial API with parallel semantics
but with minimum change of function names and argument lists. 

However, there are also limitations in parallel netCDF. Unlike HDF5, netCDF does not support 
hierarchical group based organization of data objects and since it lays out the data in a linear
order, adding fixed-size array or extending the file header may be very costly once the file
is created and has existing data stored, though moving the existing data to extended area is 
performed in parallel. Also, parallel netCDF does not provide functionality to combine two or more
files in memory through software mounting, as HDF5 does. Nor does netCDF support data compression
within its file format. Fortunately, these features can all be achieved by external software,
sacrificing some manageability of the files.

\section{Performance Evaluation}
\label{sec:results}

\begin{figure}
\begin{center}
\includegraphics[width=0.48\textwidth]{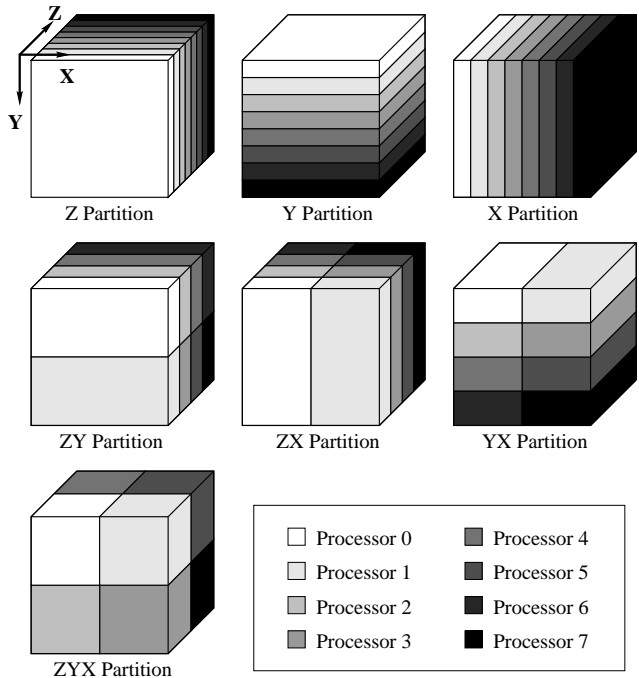}
\end{center}
\vskip -0.1in \caption{Various 3-D array partitions on 8 processors} 
\label{figure:partition}
\end{figure}

To evaluate the performance and scalability of our parallel netCDF with that of serial netCDF, 
we ran some experiments and compared the results. We
also compared the performance of parallel netCDF with that of parallel HDF5, using the FLASH I/O
benchmark.

The experiments were run on an IBM SP-2 machine. This system is a teraflop-scale 
clustered SMP with 144 compute nodes. Each compute node has 4
GB of memory shared among its eight 375 MHz Power3 processors. All the compute nodes are
interconnected by switches and also connected via switches to the multiple I/O
nodes running the GPFS parallel file system. There are 12 I/O nodes, each with dual 222 MHz
processes. The aggregate disk space is 5 TB and the peak I/O bandwidth is 1.5 GB/s.

\subsection{Scalability Analysis}

We wrote a test code (in C language) to evaluate the performance of the current
implementation of parallel netCDF.
This test code was originally developed in Fortran
by Woo-sun Yang and Chris Ding at Lawrence Berkeley National Laboratory (LBL). Basically it
reads/writes a three-dimensional array field tt(Z,Y,X) from/into a
single netCDF file, where Z=level is the most significant
dimension and X=longitude is the least
significant dimension. The test code partitions the three dimensional array 
along Z, Y, X, ZY, ZX, YX, and ZYX axes, respectively, as illustrated in
Figure~\ref{figure:partition}. All data I/O operations in these
tests used collective I/O. For comparison purpose, we prepared the same
test using the original serial netCDF API and ran it in serial
mode, in which a single processor reads/writes the whole array. 

\begin{figure*}
\begin{center}
\includegraphics[width=0.90\textwidth]{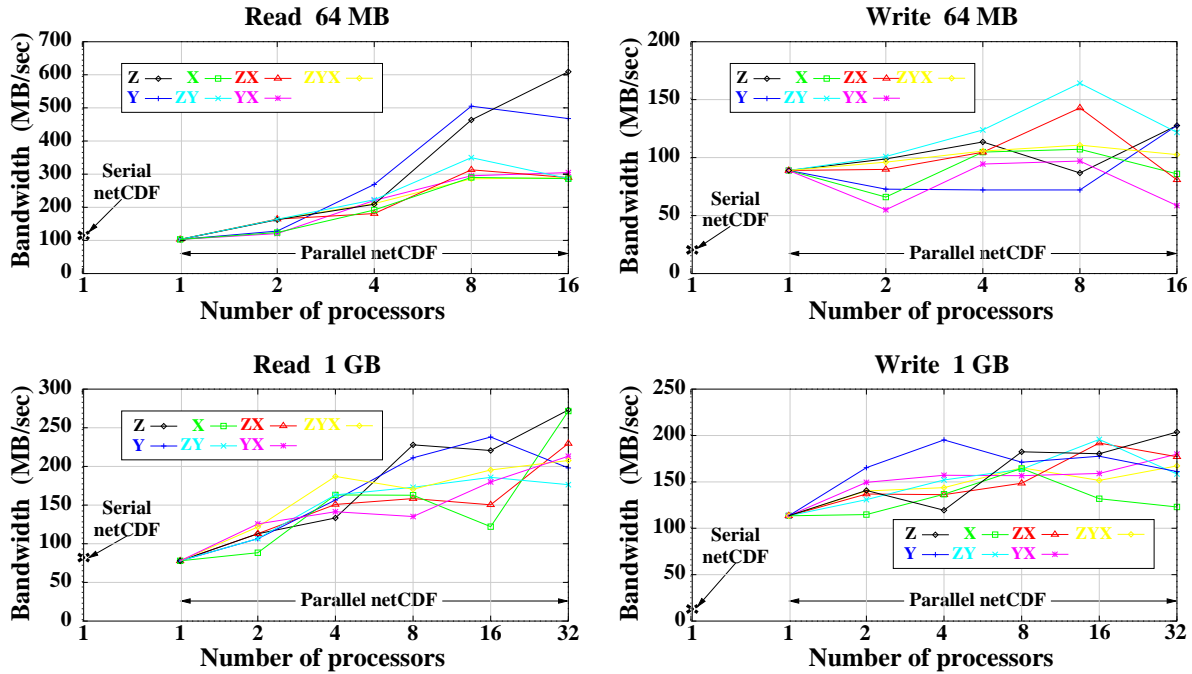}
\end{center}
\vskip -0.1in \caption{Serial and parallel netCDF performance for 64 MB and 1 GB datasets.
The first column of each chart shows the I/O performance of reading/writing the whole array 
through a single processor using serial netCDF; 
the rest of the columns show the results using parallel netCDF.
} 
\label{figure:lblbenchnetcdf}
\end{figure*}

Figure~\ref{figure:lblbenchnetcdf} shows the performance results for reading and writing 64 MB and 1
GB netCDF datasets. Generally, the parallel netCDF performance scales with the number of processes. Because of
collective I/O optimization, the performance difference made by various access patterns is small,
although partitioning in the Z dimension generally performs better than in the X
dimension because of the different access contiguity. The overhead involved is inter-process
communication, which is negligible comparing to the disk I/O when using large file size.
The I/O bandwidth does not scale in direct proportion because the number of I/O nodes (and disks) is fixed so 
that the dominating disk access time at I/O nodes is almost fixed.
As expected, the parallel netCDF outperforms the original serial netCDF 
as the number of processes increases. 
The difference between the serial netCDF performance and the parallel netCDF performance with single processor is
because of their different I/O implementations and different I/O caching/buffering strategies.
In the serial netCDF case, if, as in Figure~\ref{figure:parallelization}(a), 
multi-processors were used and the ROOT processor
needed to collect partitioned data and then perform the serial netCDF I/O,
the performance would be much worse and decrease with the number of processors
because of the additional communication cost and 
division of a large I/O request into a series of small requests.

\subsection{FLASH I/O Performance}

\begin{figure*}
\begin{center}
\includegraphics[width=0.90\textwidth]{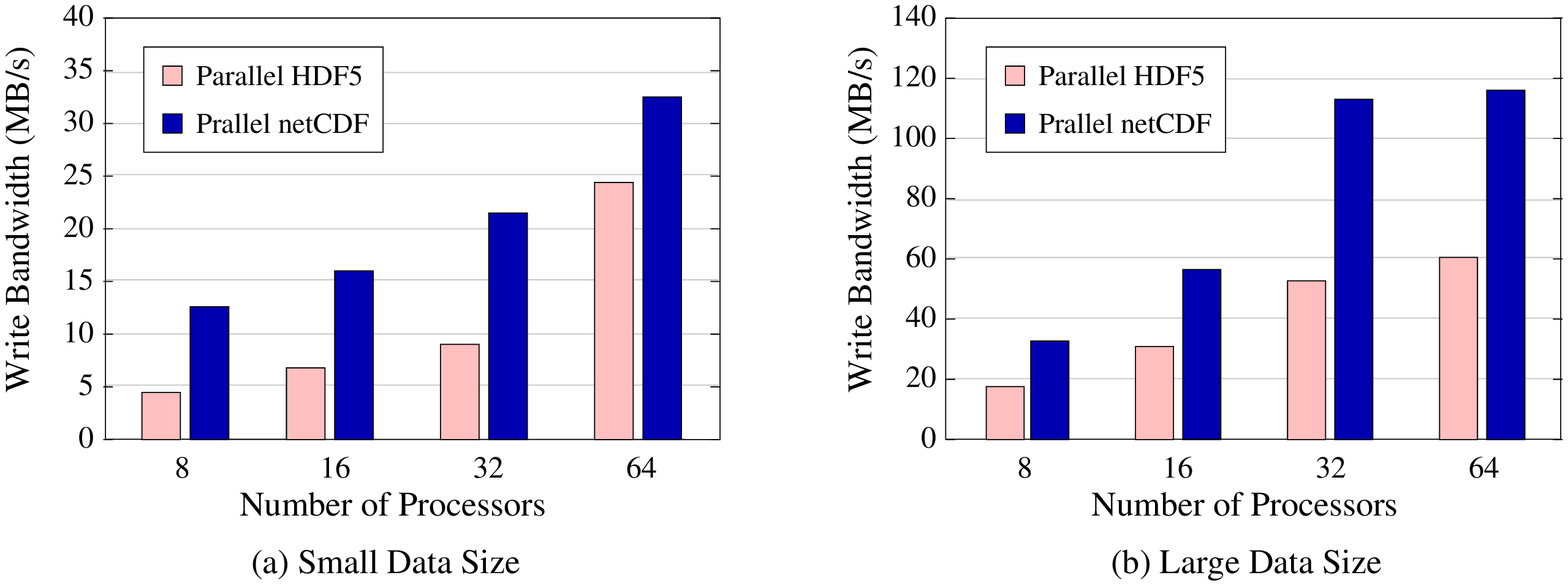}
\end{center}
\vskip -0.1in \caption{Performance of FLASH I/O benchmark using
parallel HDF5 and parallel netCDF. The two experiments use different
parameters so that the file sizes are different. Also the file
sizes are varies with the number of processors. The I/O amount
is 3MB * Number of Processors in (a), and 24MB * Number of Processors in (b).} 
\label{figure:flashio}
\end{figure*}

The FLASH I/O benchmark simulates the I/O pattern of an important
scientific application called FLASH \cite{FORT00}. It recreates
the primary data structures in the FLASH code and produces a
checkpoint file, a plotfile with centered data, and a plotfile
with corner data, using parallel HDF5. Basically, these three
output files contains a series of multidimensional arrays, and the access
pattern is simple (Block, *, ...), which is similar to the Z partition
in Figure~\ref{figure:partition}. In each of the files, the benchmark writes
the related arrays in a fixed order from contiguous user buffers, respectively.
The I/O routines in the benchmark are identical to the routines used by FLASH, so any
performance improvements made to the benchmark program will be
shared by FLASH. In our experiments, in order to focus on the data I/O performance,
we modified this benchmark, removed the part of code writing attributes,
ported it to parallel netCDF, and observed the 
effect of our new parallel I/O approach. 

Figure~\ref{figure:flashio} shows the performance results of the FLASH
I/O benchmark using parallel netCDF and parallel HDF5. 
We tested both small data size and large data size.
The parameters used in these two experiments are: 
(a) nxb = nyb = nzb = 8, nguard = 4, number of blocks = 80, and nvar = 24;
(b) nxb = nyb = nzb = 16, nguard = 8, number of blocks = 80, and nvar = 24.
Although both I/O libraries are built above MPI-IO, the
parallel netCDF has much less overhead and outperforms parallel
HDF5 by almost doubling the overall I/O rate. The extra overhead
involved in the current release of HDF5 (version 5-1.4.3) includes
inter-process synchronizations and file header access performed internally in parallel open/close of
every dataset (analogous to a netCDF variable) and recursive handling
of the hyperslab used for parallel access, which makes the packing of
the hyperslabs into contiguous buffers take a relatively long time.

\section{Conclusion and Future Work}
\label{sec:conclusion}

In this work, we extend the serial netCDF interface to facilitate parallel access, and we provide an
implementation for a subset of this new parallel netCDF interface. By building on top of MPI-IO, we
gain a number of interface advantages and performance optimizations users can benefit from
by using this parallel netCDF package, as shown by our test results. So far, 
a number of users from LBL, ORNL, and University of Chicago are using our parallel netCDF library.

Future work involves developing a production-quality parallel netCDF API (for C, C++, Fortran, and
other programming languages) and making it freely available to the high-performance computing community.
Moreover, we need to develop a mechanism for matching the file organization to access patterns, and
we need to develop cross-file optimizations for addressing common data access patterns.

\subsection*{Acknowledgements}


This work is sponsored by Scientific Data Management Center of DOE SciDAC ISICs program and jointly
conducted at Northwestern University and Argonne National Laboratory. This research was also
supported in part by NSF cooperative agreement ACI-9619020 through computing resources provided by
the National Partnership for Advanced Computational Infrastructure at the San Diego Supercomputer
Center.

We thank Woo-Sun Yang from LBL for providing us the test code for
performance evaluation  and Nagiza F. Samatova and David Bauer at
ORNL for using our library and for giving us feedback and
valuable suggestions.


\end{document}